# Programmable and nonvolatile computing with composition tuning in thin film lithium niobate

Abhiram Devata[1,*], Axel Magaña Ponce[2], David Barton[1,*]

[1] Department of Materials Science and Engineering, Northwestern University, 2220 Campus Drive Evanston, IL 60208, USA

[2] Department of Physics, Elmhurst University, 190 S Prospect Avenue, Elmhurst, IL 60126, USA

[*] Corresponding Authors: adevata@u.northwestern.edu, dbarton@northwestern.edu

## Abstract

Matrix-vector multiplications are fundamental operations in artificial intelligence and high-throughput computations, and are executed repeatedly during training and inference. Their high energy cost in electronic processors motivate scalable photonic computing approaches that reduce the energy required per operation. Thin film lithium niobate is a dominant photonic platform due to its large electro-optic effect. However, it lacks nonvolatile index tuning mechanisms, which promise to pave the way for energy-efficient photonic computing. Here, we explore electrochemical lithiation as a route to nonvolatile matrix-vector multiplications in thin film lithium niobate. The lithium niobate phase is stable at room temperature over a 2% Li composition window with an associated composition-dependent refractive index. We computationally demonstrate this as a programmable, low-loss approach to perform matrix-vector multiplications by using composition to control matrix weights. We design Mach-Zehnder interferometers to perform image processing tasks under realistic material loss constraints. We also design microring resonators for iterative weight updates, using gradient descent training to program target matrix operations with matrix-vector multiplication accuracy validated at 1.6% average relative error. These demonstrations show a facile route towards nonvolatile photonic computing in thin film lithium niobate, addressing a critical requirement for energy-efficient photonic matrix operations at scale.

## 1. Introduction

Developing energy-efficient hardware for artificial intelligence (AI) is critical as the energy consumption of data-intensive workloads continues to grow.[1–4] Matrix-vector multiplications (MVMs) form the backbone of many AI algorithms, where they are performed repeatedly during both training and inference.[5,6] In conventional electronic hardware, the switching energy cost and data movement overhead of each MVM accumulates across these repeated operations.[7,8] Electronic in-memory computing architectures address part of this problem by co-locating computation and memory to reduce data movement.[9,10] Integrated photonics offer complementary advantages by performing MVMs in the optical domain, reducing both data movement and switching energy compared to electronic processors.[6] However, photonic implementations using active tuning mechanisms require continuous power to maintain programmed weights. Nonvolatile photonic computing promises to eliminate this static power requirement, with potential for compounded energy savings across repeated operations.[11]

Among photonic platforms, thin film lithium niobate (TFLN) offers advantages for scalable integrated photonics due to its excellent optical properties, including a large electro-optic effect ($r_{33}$ = 30 pm/V), low waveguide propagation loss ($\alpha$ = 1.3 dB/m), and wafer-scale processing.[12–15] These properties enable compact, high-performance modulators that have recently been demonstrated for MVMs.[16–20] However, TFLN currently lacks a nonvolatile index tuning mechanism. While TFLN's photorefractive effect enables semi-nonvolatile tuning, it is quite weak (mode index change of $4\times10^{-5}$) and fully relaxes in less than a day, making it incompatible with weight storage for MVMs.[21] This limits MVM demonstrations on TFLN to active tuning that requires continuous power to maintain matrix weights.

Phase change materials provide nonvolatile operation on other platforms through reversible switching between crystalline and amorphous phases. This changes the real and imaginary components of the material and thus mode index, thus representing MVM weights.[11,22–28] However, efforts to integrate phase change materials with integrated photonics have focused primarily on silicon platforms. The complexity of co-fabricating phase change material switching methods onto low-loss TFLN circuits is primarily driven by TFLN's insulating nature, reducing the conductivity required for integrated microheaters to toggle the phase change material.[29,30] TFLN's large electro-optic bandwidth is already well suited for high-throughput data modulation, while nonvolatile MVM processors on the same platform would enable persistent weights and reduce energy consumption.[18,31] Enabling nonvolatility on TFLN will unify existing high-speed modulators with low-power inference architecture to monolithically process large batches of data.

Bulk studies in lithium niobate show that composition can be a potential tuning knob for optical properties. Phase diagrams predict a pure lithium niobate phase with a 2% composition window at room temperature starting from the congruent composition.[32,33] Over this window, a refractive index change of +0.002 and -0.016 for ordinary and extraordinary axes respectively is primarily driven by vacancy chemistry and its anisotropic contribution to lithium niobate's electron polarizability.[34,35] Moreover, lithiating LN from its congruent composition by filling vacancies is expected to blueshift the absorption spectra by reducing lattice deformation.[36,37] This compositionally tunable index change without significant increases to absorption is promising for designing scalable photonic circuits. Since TFLN devices originate from bulk crystals grown at the congruent composition, developing methods to controllably tune TFLN composition post-growth is of interest for nonvolatile refractive index control.

Here we explore how electrochemistry can be used to tune composition in TFLN single crystals for nonvolatile photonic computing structures. We exploit TFLN as an electrolyte through which Li can be added and removed, much like in a battery. Amorphous lithium niobate has demonstrated success as a cyclable electrolyte for electrochromic and plasmochromic devices.[38,39] The resulting composition-dependent waveguide mode index change is sufficiently large (0.008) to design and simulate interferometers and ring resonators amenable to computing. We then show the potential of this platform by cascading and reprogramming these components to perform MVM operations such as image processing and neural network training prerequisites, establishing a practical pathway to nonvolatile photonic computing in TFLN.

## 2. Results and Discussion

### 2.1. Working Principle of Composition Tuning

The proposed structure of our compositionally tunable TFLN unit is shown in Figure 1a. We consider monolithically etched rib waveguides on 600 nm TFLN with an $SiO_2$ buried oxide. A metal back contact is included beneath the buried oxide to prevent absorption-induced propagation loss of guided modes.[40–42] A polymer electrolyte "cladding" and Li source are included above the TFLN.[43] Once the TFLN waveguides are etched using a standard protocol, the electrolyte can be spin-coated on and patterned with photolithography. The Li film can be deposited with electron-beam evaporation and patterned using techniques designed for microbatteries.[44–48] The simulated $TE_{00}$ mode for unlithiated TFLN at a wavelength of 1550 nm is included inset in Figure 1a.

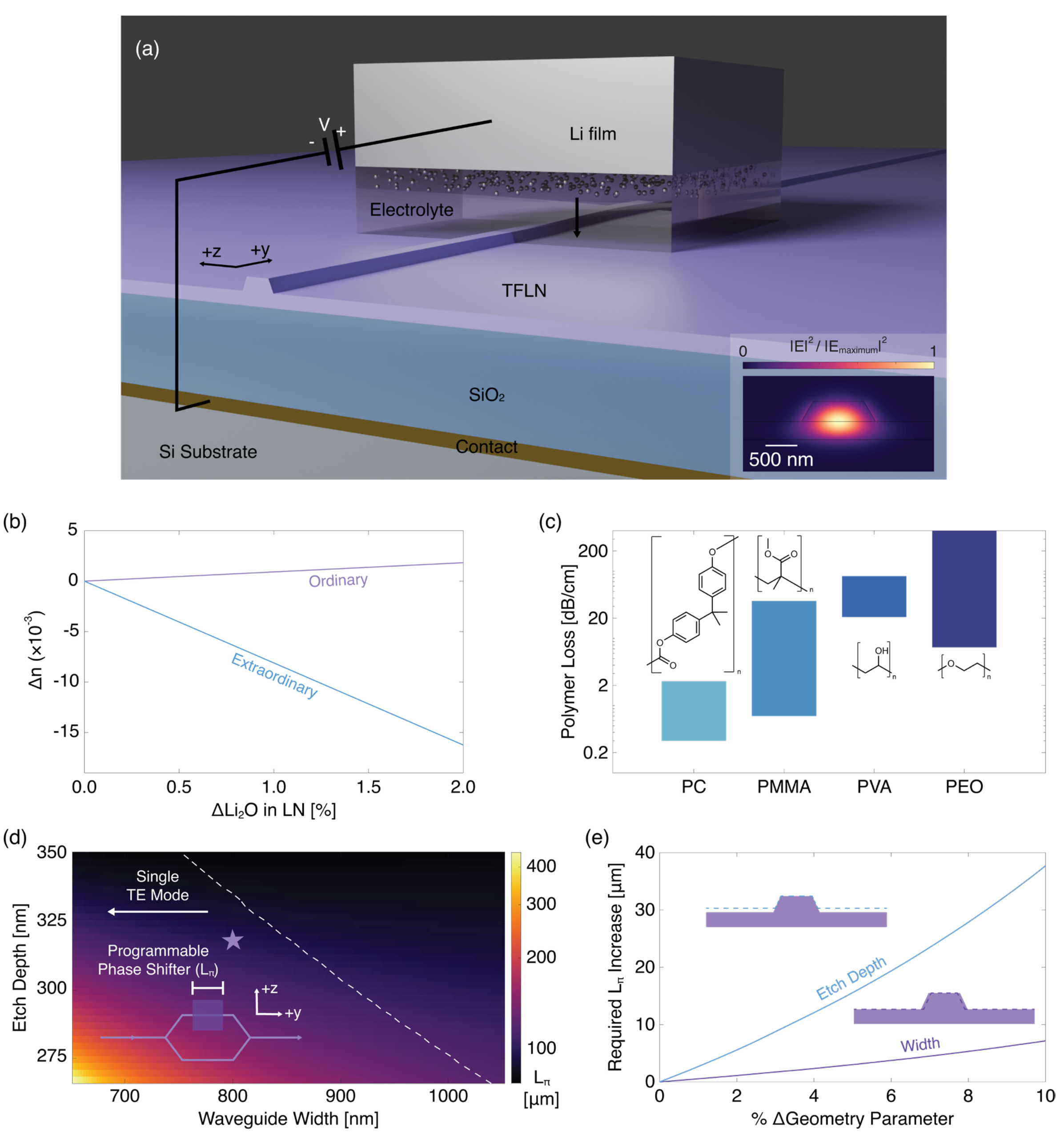

**Figure 1.** Schematic of TFLN waveguide lithiation and interferometer design. (a) Nonvolatile TFLN stack being lithiated. A positive bias with respect to the bottom electrode drives Li+ ions into etched structures on TFLN, resulting in a change in refractive index. (a, inset) Simulated $TE_{00}$ mode profile of unlithiated stack. (b) Bulk ordinary (purple) and extraordinary (blue) refractive indices at 1550 nm change with Li concentration. Adapted from [34]. (c) Optical loss from poly(carbonate) (PC)[49–51], poly(methyl methacrylate) (PMMA)[52–54], poly(vinyl alcohol) (PVA)[55–57], and poly(ethylene oxide) (PEO)[58–60], which can be used as matrices for Li electrolytes. (d) $L_\pi$ required for MZI to sweep output response over 2% composition window. The white dashed line represents the single $TE_{00}$ mode regime. (e) Additional length required

to correct waveguide fabrication errors in etch depth (blue) and width (purple) via nonvolatile optical trimming.

When a positive bias is applied to the system, Li ions are driven into the TFLN, changing both the ordinary and extraordinary refractive indices. We restrict our simulations to a 2% composition window starting from congruent LN to eliminate the possibility of a solid phase transition.[32,33] Figure 1b reproduces experimentally measured index changes at 1550 nm with increasing Li concentration, showing a nearly linear relation for added Li across the full composition window.[34] Moreover, Figure 1b highlights the anisotropy of adding Li to LN, as the absolute change in extraordinary index (-0.016) over the composition window is eightfold the ordinary index change (+0.002). By comparison, TFLN electro-optic modulators have demonstrated index changes of up to 0.003, while most TFLN electro-optic modulators produce index changes of around 0.0001.[12,13,16] Nonvolatile weight programming is expected to take a few seconds to fully (de)lithiate 600 nm TFLN over the 2% composition window, but elevated temperatures can increase the (de)lithiation speed.[39,61,62]

Waveguides were designed with a 60 degree sidewall angle.[12,14,63,64] A refractive index of 1.486 was used for the electrolyte to model mode confinement.[65,66] Organic polymers typically used for solid-state Li electrolytes are absorbing in the near infrared regime due to vibrational modes of the C-O bond.[67,68] While highly dependent on phase and molecular weight, this additional propagation loss from the polymer electrolyte $\alpha_{PE}$ is accounted for in subsequent sections.[65,66,69] Solution-casted formulations of poly(carbonate), poly(methyl methacrylate), and poly(ethylene oxide) have losses ranging from roughly 0.5 to 20 dB/cm at 1550 nm as shown in Figure 1c, and their absorption decreases with increasing molecular weight.[49–60] These polymers can serve as host matrices for powder electrolytes without significant increases to absorption. While polymer claddings have already demonstrated successful integration on TFLN, if $\alpha_{PE}$ severely restricts the available electrolyte claddings, these computing systems can be redesigned for the O-band centered at 1310 nm, where organic vibrational modes are expected to be less active.[53,67,68,70,71] Moreover, mild fluorination of these polymers can further blueshift their absorption spectra by replacing lossy C-OH C=O overtones with higher energy C-F bonds, allowing for operation at 1550 nm with reduced ionic conductivity and thus slower switching speeds.[72–74]

## 2.2. Mach-Zehnder Interferometers for Image Processing

A refractive index change from lithiation allows for relative propagating phase accumulation and thus interferometry with Mach-Zehnder Interferometers (MZIs). The critical design feature for an MZI (Figure 1d, inset) is its π-phase shift length $L_\pi$, which depends on the difference in mode index between the two arms according to

$$L_\pi = \frac{1}{2}\left|\frac{\lambda_0}{\Delta n_{eff}}\right| = \frac{1}{2}\left|\frac{\lambda_0}{n_{lithiated} - n_{unlithiated}}\right| \quad (1)$$

where $\lambda_0$ is the free space wavelength (1550 nm), $n_{lithiated}$ and $n_{unlithiated}$ are the effective $TE_{00}$ mode indices for the lithiated and unlithiated waveguides respectively, and $\Delta n_{eff}$ is the difference between lithiated and unlithiated effective $TE_{00}$ mode indices. Depending on the waveguide's width and etch depth, $\Delta n_{eff}$ ranges from $8.6\times10^{-4}$ to $5.5\times10^{-3}$.

We design MZIs in X-cut TFLN (z axis oriented perpendicular to MZI arms and in plane of TFLN layer) using TE modes to maximize the index modulation, resulting in a smaller device footprint. Aligning a majority of the mode's electric field with the crystal axis that experiences a larger index change upon lithiation will increase the change in effective mode index, thus reducing the corresponding $L_\pi$. Figure 1d shows that $L_\pi$ decreases for the $TE_{00}$ mode as waveguide width and etch depth increase. As either parameter increases, more of the mode overlaps with the larger rib waveguides, thus increasing the portion of the mode that experiences the index change from lithiation. As a result, $\Delta n_{eff}$ increases, decreasing the required length for a π-phase shift. We additionally constrain our MZI design to single $TE_{00}$ mode waveguides, as marked to the left of the dashed line in Figure 1d. Based on these constraints, we choose a waveguide width of 800 nm and etch depth of 315 nm for MZIs as indicated by the star on Figure 1d, yielding a $\Delta n_{eff}$ of 0.0078 and an $L_\pi$ of 100 μm. This length is comparable to thermo-optic[75–77] and carrier accumulation[78–80] MZIs on the silicon on insulator platform, and is three to fifteen times shorter than carrier depletion[81–83] MZIs. It is also around two orders of magnitude shorter than conventional TFLN electro-optic MZIs.[17,84–86]

We can additionally utilize this effect for optical trimming, as depicted in Figure 1e. Using the chosen waveguide geometry mentioned above, if either waveguide geometry parameter decreases by up to 10% from the chosen geometry in Figure 1d, no more than 40 μm of additional length would be required to ensure a full π-phase shift is still possible. Since smaller rib waveguides experience lower $\Delta n_{eff}$, they would require a longer $L_\pi$ to compensate for the insufficient phase accumulation between the MZI's arms. Larger rib geometries do not require additional length as they would experience $\Delta n_{eff}$ larger than the target from lithiation. To correct for large fabrication errors, a second phase shifter in the same arm with length according to Figure 1e can serve as a one-time nonvolatile phase correction to ensure the full $\pi$-phase shift is achievable. Alternatively, a single phase shifter with length equal to the sum of $L_\pi$ and the length from Figure 1e can be tuned over a smaller composition range. Both options are viable calibration methods that require balancing energy consumption with design simplicity. A tighter fabrication tolerance will reduce this length.

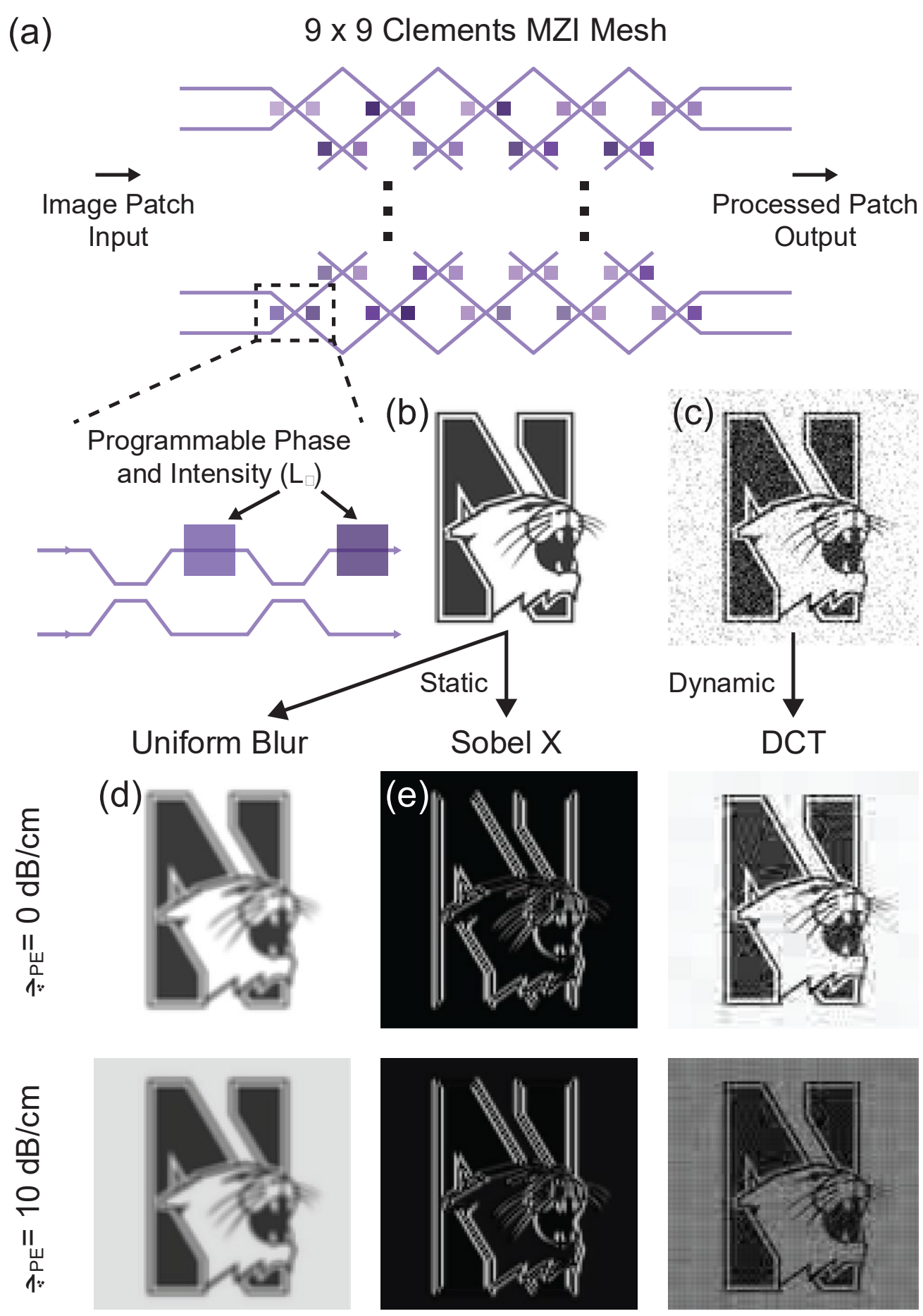

**Figure 2.** Demonstration of lithiated MZIs for image processing. (a) 2×2 MZI mesh with 2 independent phase shifters of length $L_\pi$ are used to build larger MZI meshes to perform image processing MVMs. (b) Mascot used for static image processing via kernel operations. (c) Mascot with artificially added Gaussian noise with $\sigma = 0.1$. (d) Simulated uniform blur kernel processing with $\alpha_{PE} = 0$ dB/cm (top) and $\alpha_{PE} = 10$ dB/cm (bottom). (e) Simulated vertical edge detection kernel processing with $\alpha_{PE} = 0$ dB/cm (top) and $\alpha_{PE} = 10$ dB/cm (bottom). Mesh is tolerant to 10 dB/cm of electrolyte loss, although features lose resolution and intensity. (f) Simulated discrete cosine transform on artificially noised image with no propagation loss from electrolyte (top) and 10 dB/cm electrolyte propagation loss (bottom).

By expanding this design towards 2×2 MZIs with 2 independently programmable phase shifters as shown at the bottom of Figure 2a, we can represent any real 2×2 unitary matrix.[87] We utilize these 2×2 bocks via an interferometer mesh capable of performing MVMs. To represent arbitrary real matrices that are used to process a desired image, we employ singular value decomposition to deconstruct a target matrix into unitary and diagonal matrices.[88] The Clements arrangement of 2×2 MZIs shown in Figure 2a describes a footprint-efficient way to represent unitary matrices of arbitrary size.[89] Similarly, diagonal matrices can be represented by a row of phase shifters as none of the output rows are affected by inputs from other rows. This architecture allows phase shifters to perform real-valued MVMs. We specifically use a 9×9 Clements mesh to perform computations on multiple unraveled 3×3 matrices in a single pass, as each 3×3 matrix represents a full row within the mesh. Pixel intensities of the images in Figures 2b and 2c are fed into the left side of the mesh, with the right side outputting the processed images of Figures 2d, 2e, and 2f.

We first discuss static image processing by computing 4 3×3 kernels simultaneously: Sobel X (vertical edge detection), Sobel Y (horizontal edge detection), Laplacian (sharpening), and uniform blur (blurring).[90–92] Figure 2b shows the initial 100×100 pixel squared grayscale image that the kernel processing MVM is performed on. One of the key factors dictating viability of lithiation as a nonvolatile MVM platform is the propagation loss due to a polymer electrolyte αPE that is required to enable lithiation. The uniform blur and Sobel X results of this MVM performed on Figure 2b are shown in Figures 2d and 2e for two different $\alpha_{PE}$ values. The primary difference between the two losses of the uniform blur kernel in Figure 2d is that white regions such as the background are 13% dimmer for higher propagation loss. If these were normalized to take up the full intensity range (0 to 1), noise would also be amplified more strongly for lower contrast images, which is undesired. This effect is also observed in the Sobel X kernel results in Figure 2e, although the difference is less substantial between the two $\alpha_{PE}$ values as the higher intensity regions are smaller. The intensities for edge pixels are 15% brighter in the lossless simulation, and finer features such as teeth and whiskers are more clearly resolved. $\alpha_{PE}$ effectively attenuates the intensity of the gradient kernel, resulting in weaker edges. Even with 10 dB/cm of propagation loss, edges are still clearly identifiable across the entire image. As such, within this range of polymer loss, we can passively operate the 9×9 MZI mesh to simultaneously compute up to 9 kernels on large sets of images.

Single pass static operation limits the types of image processing algorithms that can be run on our 9×9 MZI mesh to 3×3 kernels like those used for Figures 2d and 2e. To expand beyond this, we next discuss dynamic image processing via the discrete cosine transform (DCT) algorithm on the same mesh.[93] This algorithm is performed on Figure 2c, which is identical to Figure 2b with the exception of artificially added gaussian noise ($\sigma = 0.1$). The results of the algorithm with a 9×9 DCT matrix for different $\alpha_{PE}$ are shown in Figure 2f. Four passes are required to ensure both row and column data are accounted for in both the forward and inverse computation. After the first MVM is finished, not only are the results are fed back into the MVM in a different orientation, but a filtering step is also required before weights are reconfigured for the inverse computation, making this a dynamic process with weights that must update in real-time based on the current step of the algorithm. In a lossless mesh, almost all of the noise is successfully removed from the image, resulting in a cleaner image than the input. Patches that include finer features such as the mascot's whiskers and fur have lost some of their sharpness. However, when loss is added, the mesh performs substantially worse, with some of the coarser features such as the upper slant in the "N" also not being fully resolved. Due to intensity rescaling used during postprocessing to account for signal variation from $\alpha_{PE}$, lower contrast features such as those around the mouth are washed out. The multi-pass process results in loss having a larger impact on the output image compared to the static image processing. Nevertheless, Figure 2f shows that DCT and other multi-pass image processing algorithms are viable on this hardware platform even in the presence of loss. Therefore, the same 9×9 MZI hardware can be electrochemically reprogrammed to perform a wide range of image processing tasks. Moreover, for batch processing many and/or large images, weights do not need to be reset, further improving system efficiency due to the nonvolatile nature of the weights.

### 2.3. Microring Resonators for Neural Networks

Microring resonators (MRRs) are another major class of photonic devices that have demonstrated success at performing MVMs on the silicon-on-insulator platform.[94] Figure 3a depicts a crossbar MRR typically used for these operations, where each resonator can act as a weight in a matrix. Here, we design crossbar MRRs for the TFLN platform using lithiation as the weight-setting mechanism. Waveguide crossings have already been designed in TFLN.[95] For simulation simplicity, we explore Z-cut TFLN (z axis oriented perpendicular to plane of TFLN layer) MRRs, as X-cut designs are susceptible to polarization conversion due to the crystal's in-plane optical anisotropy.[96,97] Additionally, we choose to lithiate only a quarter of the ring opposite to the waveguide crossing, as shown in Figure 3a. This ensures lithiation doesn't change the coupling coefficient. We use MRRs with a radius of 60 μm to minimize footprint without being dominated by bending loss, which provides fabrication flexibility. Additionally, we use an etch depth of 300 nm.[84] The resulting bent $TE_{00}$ optical mode in the resonator is shown inset in Figure 3a.

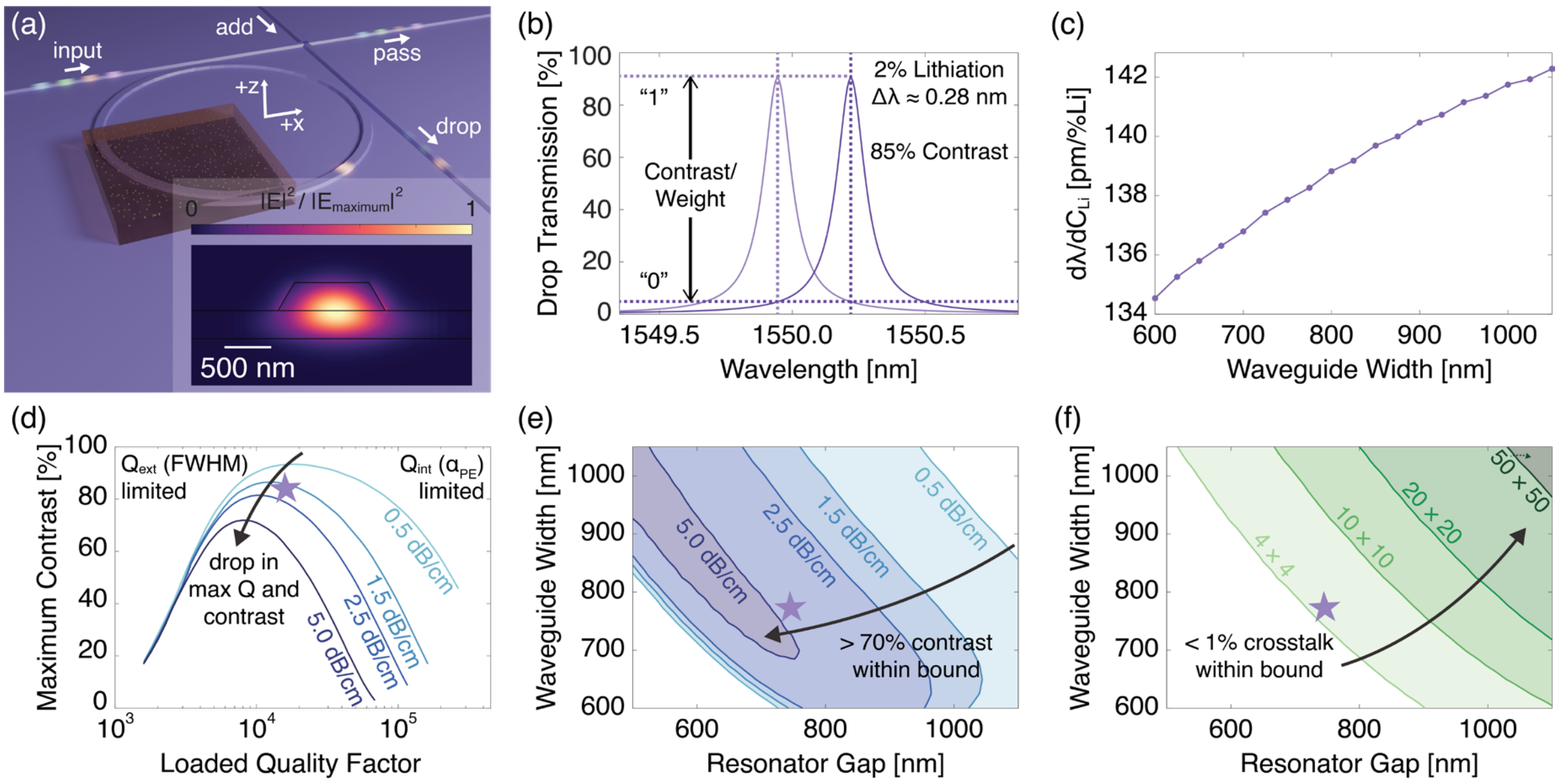

**Figure 3.** Design parameters and space for computing with TFLN crossbar MRRs. (a) Lithiating a quarter of the ring in a crossbar MRR sets a weight that determines the output intensity in the drop port. (a, inset) Simulated mode profile in a 60 μm radius resonator showing strong light confinement. (b) Contrast is defined as the difference in transmission between the unlithiated (lighter curve) and lithiated (darker curve) states at the resonant wavelength of the unlithiated state. (c) Shift in resonant peak position with respect to lithiation as a function of waveguide width. (d) Maximum contrast possible versus loaded MRR quality factor decreases as $\alpha_{PE}$ increases. (e) 70% contrast contours over the resonator gap and waveguide width design space decrease with increasing $\alpha_{PE}$. The line is included as a guide to the eye. (f) Minimum geometry parameters required for 1% crosstalk increase as the desired mesh size increases. The star in (d), (e), and (f) marks the optimized MRR design. The contours in (d), (e), and (f) represent discrete values labeled on the plot, with darker colors indicating larger values.

The key parameters to design for in our MRR crossbar units are contrast, tuning efficiency, and crosstalk. The contrast, defined in Figure 3b, is the maximum difference in drop port transmission after lithiation. It determines the maximum programmable weight without amplification and therefore sets the effective bit depth for MVM operations. Reliable weight discrimination across the full transmission range requires high contrast, which depends on the balance between coupling strength and propagation loss, both of which are determined by waveguide geometry and material properties. The tuning efficiency, shown in Figure 3c, quantifies the total change in resonance position divided by the total change in Li concentration.

This is primarily influenced by waveguide geometry, as an increase in width increases the mode index change from lithiation due to a larger field overlap with the TFLN. As a result, higher tuning efficiency enables larger resonance shifts for the same lithiation range, directly increasing the available contrast. At a first glance, this suggests that wider waveguide widths are preferable. However, the geometric parameters that determine tuning efficiency simultaneously control the loaded quality factor in combination with propagation loss. Since contrast depends on the loaded quality factor, optimizing for high tuning efficiency introduces competing constraints that limit the resulting contrast.

Figure 3d reveals these trade-offs by identifying the quality factor with maximum contrast for different propagation losses. Higher loaded quality factors, dominated by the MRR's intrinsic quality factor ($Q_{int}$), are characterized by limited coupling into and out of the resonator which reduces the extinction ratio, resulting in lower contrast despite the lower full width at half maximum (FWHM) of the resonant peaks. For significantly lower quality factors dominated by the MRR's extrinsic quality factor ($Q_{ext}$), the broadened resonance limits contrast and increases spectral overlap between adjacent channels, leading to increased crosstalk. For our system, $\alpha_{PE}$ is the dominant intrinsic loss, as state-of-the-art TFLN fabrication yields propagation losses of 1.3 dB/m.[14,63] Therefore, increasing $\alpha_{PE}$ reduces the quality factor at which maximum contrast is achieved. The optimal MRR design therefore balances FWHM with $\alpha_{PE}$ to maximize contrast while limiting crosstalk.

Mapping these quality factor trade-offs back onto MRR geometry reveals how intrinsic and extrinsic losses constrain the achievable contrast. Figure 3e highlights the decrease in MRR geometries available that satisfy a 70% contrast threshold for increasing $\alpha_{PE}$. Geometries with weaker mode confinement and stronger evanescent coupling are more tolerant to propagation loss as these contours shown in the bottom left of Figure 3e are bunched together. However, their broader resonance FWHMs increase spectral overlap between channels, which impacts crosstalk. As the mode confinement improves, the system becomes intrinsically dominated, making contrast highly sensitive to $\alpha_{PE}$. Together, these two trends establish $\alpha_{PE}$ as a primary design constraint that determines the feasible balance between contrast, crosstalk, and geometry.

For these MRR crossbars to be cascaded into an array for MVM operations, the crosstalk between adjacent MRRs must be minimized without sacrificing channel bandwidth. Depending on the desired matrix size and maximum acceptable crosstalk, a lower bound on the resonator gap and waveguide width will satisfy this threshold. Figure 3f demonstrates this lower geometry bound for $\alpha_{PE}$ = 1.5 dB/cm and a 1% crosstalk threshold, which depends on the channel spacing, tuning efficiency, and resonance FWHM.[98] Since the effective mode index varies weakly with waveguide width, the free spectral range (FSR) is approximately constant for all simulated geometries. The target matrix size therefore sets the required channel bandwidth given by the FSR divided by the larger matrix dimension, with the optimal lithiation-induced resonance shift chosen as half the channel bandwidth to maximize contrast while minimizing spectral leakage. Higher quality factors reduce crosstalk as narrower peaks allow for more channels to fit within a single FSR. Alternatively, reducing the ring radii to increase the FSR will also reduce the crosstalk.[99] However, high quality factors come at the cost of lower drop port transmission due to reduced coupling into and out of the MRR. To meet the crosstalk threshold for larger matrices closer to the top right of Figure 3f, the MRRs require a lower $\alpha_{PE}$ and/or a reduced contrast to open up the available geometry space. Polymer claddings on TFLN have demonstrated quality factors of $10^5$, showing promise for future scalability.[69] Given a target mesh size, desired crosstalk threshold, and electrolyte propagation loss, Figures 3d, 3e, and 3f determine an optimal MRR geometry that balances crosstalk and contrast. For a target 4×4 matrix with a crosstalk of less than 1% and $\alpha_{PE}$ = 1.5 dB/cm, we choose both resonator gap and waveguide width of 750 nm to yield a maximum contrast of 80% at a quality factor of $10^4$. This point is marked by a star in Figures 3d, 3e, and 3f.

Once a robust MRR geometry has been selected, the crossbar rings can be cascaded to create a MVM processor similar to a systolic array, as shown in Figure 4a.[100,101] Each MRR is individually programmable, allowing nonvolatile weights between 0 (fully lithiated) and 1 (unlithiated) to be set. The MRRs have varying radii over a 4 µm range according to the wavelength position of each channel such that no row or column has the same ring size multiple times. Once a specific wavelength passes into a ring it is resonant with, a fraction of the input light determined by the ring's weight exits through the drop port and travels down the column of the array, until it is collected by a photodetector.[94]

To demonstrate the performance of this system, we simulate a large number of random matrices multiplied by random vectors, with the relative error $e_{rel}$ shown in Figure 4b. We define relative error according to

$$e_{rel} = \frac{|y_{simulated} - y_{expected}|}{|y_{expected}|} \tag{2}$$

where $y_{simulated}$ and $y_{expected}$ are the simulated and ground truth output vectors. For $\alpha_{PE}$ = 1.5 dB/cm, we obtain a tight distribution centered around 1.5% error with a standard deviation of 0.7% across 10000 random simulations. The nonzero center and slight right skew arise from systematic crosstalk between nearby MRRs, where unintended sampling of Lorentzian tails of other MRRs leads to accumulated relative error. However, we retain a near-Gaussian error distribution, meaning that our MRR mesh can perform MVMs across the full range of weights with high accuracy and precision. Thus our MRR design can still reliably perform MVMs even in the presence of loss. This metric is on par with 1-2% relative error obtained by other photonic MVM processors.[102–104] While operating each MRR in the mesh sequentially produces no crosstalk, the operational time for the full array scales as the square of the matrix size, which severely limits practical usage. Crosstalk can be further reduced by accepting lower contrast during operation. Moreover, the larger FSRs of smaller MRR radii allow for the same number of channels to be more spaced out, which reduces crosstalk.

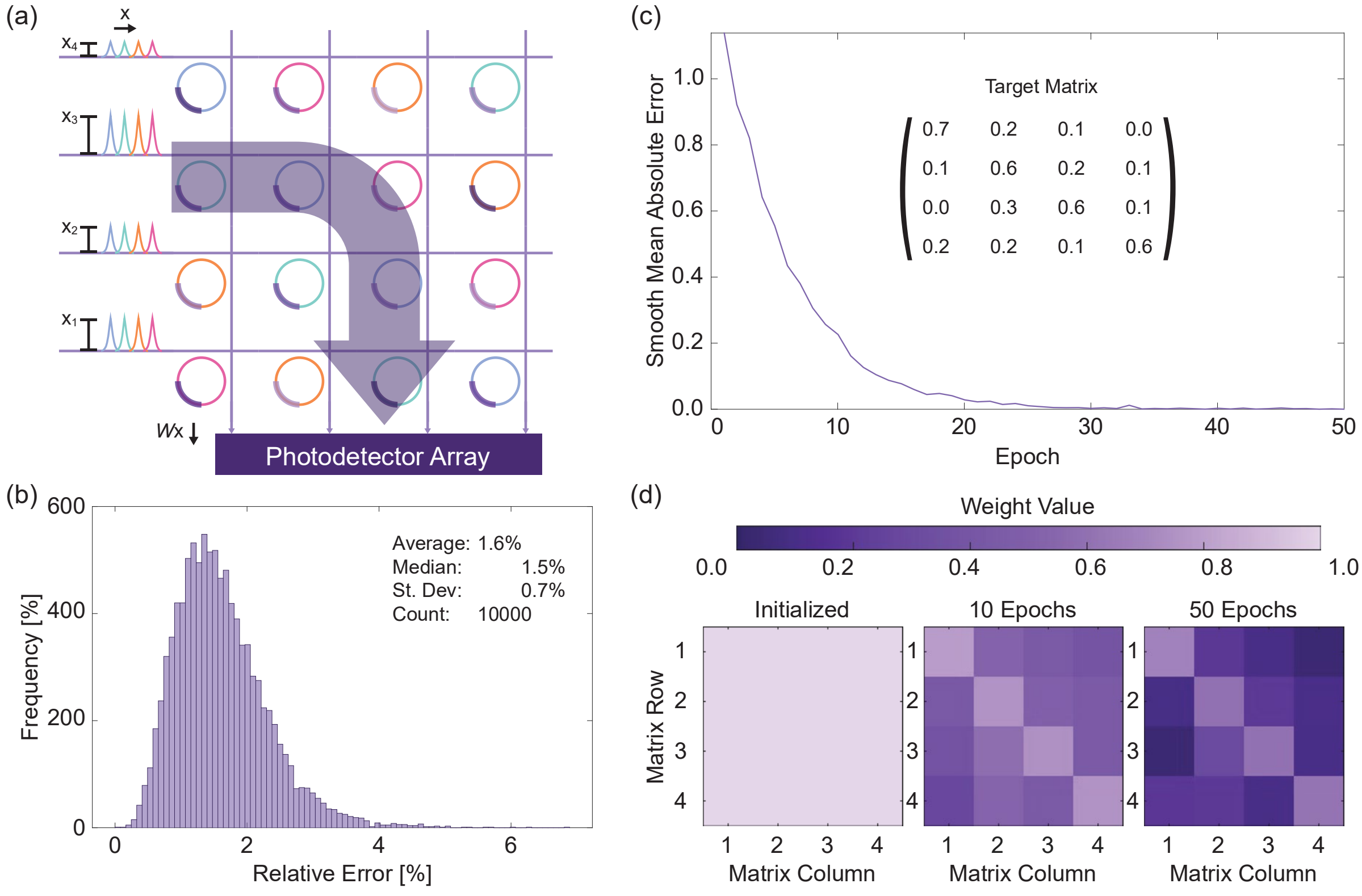

**Figure 4.** Demonstration of lithiated MRRs for matrix representation training. (a) 4×4 MRR crossbar array configuration used to compute MVMs. (b) Distribution of MVM accuracy for 10000 random matrices and vectors shows consistency in performance. (c) Stochastic gradient descent to train an unlithiated mesh to represent a target matrix converges despite propagation loss of 1.5 dB/cm and 10% noise.

We build on this reliability by training our mesh to represent a target matrix because our mesh can compute MVMs accurately. To further demonstrate its applicability for artificial intelligence hardware, we introduce random noise over the simulated resonant peak at a signal to noise ratio of 6.5 dB. Starting from a fully unlithiated system where all weights are 1, we use stochastic gradient descent to update the MRR array's weights until the desired matrix is represented. Figure 4c shows the convergence of the smooth mean absolute error of the system to represent the desired matrix via lithiated weights in our system over 50 epochs. Smooth mean absolute error was selected as the loss function as it penalizes both large and small deviations from ideality similarly.[105,106] Within 30 epochs, the target matrix is accurately represented with an error of 0.1% that is on par with other photonic MVM processor demonstrations.[104,107,108] By combining this framework with the ability for this mesh to also compute transpose matrices without requiring weight resetting, this hardware lends itself to backpropagation that uses the matrix transpose to update the weights.[94,109] In a nonvolatile architecture, this means that energy is only put into lithiating or delithiating the weights once per training cycle, instead of requiring power to constantly maintain weights for the duration of operation. Once training is finished the system acts passively to represent the target matrix.

## 3. Conclusion

We designed compact, scalable, and reliable TFLN computing structures that can host nonvolatile weights using electrochemistry. We then cascade and reprogram these components in simulation to efficiently perform meaningful MVM tasks, such as image processing and neural network training steps. These systems can function as energy efficient MVM accelerators for repeated operations, making TFLN more capable of meeting the demands of modern computing. This approach allows for the monolithic integration of electro-optic and nonvolatile index control on TFLN, providing functionality that was previously unavailable on a single platform.

## 4. Methods

Effective mode index for compositionally tuned X-cut straight waveguides were simulated using COMSOL Multiphysics with the Wave Optics module. These mode indices were used to calculate $L_\pi$ according to Equation 1. The Northwestern Wildcat Mascot was obtained from Wikimedia Commons, resized into a square, rescaled to 100 pixel × 100 pixel, and converted to grayscale.[110] Clements mesh decomposition and MZI image processing were simulated via MATLAB.

Effective mode indices for compositionally tuned Z-cut waveguides with a bending radius of 60 μm were simulated using Lumerical FDE. The coupling gap was simulated using Lumerical FDTD, with bends being approximated by cubic Bezier curves. These were used in Lumerical Interconnect to simulate the response of individual MRRs. MRR matrices used for reliability and matrix representation training were simulated via MATLAB.

## Acknowledgements

This research was supported in part through the computational resources and staff contributions provided for the Quest high performance computing facility at Northwestern University which is jointly supported by the Office of the Provost, the Office for Research, and Northwestern University Information Technology. A.D. was supported by the National Science Foundation Graduate Research Fellowship under Grant No. DGE-2234667. A.M. was supported by the Materials Initiative for Comprehensive Research Opportunity (MICRO) at NU. MICRO is generously supported by an unrestricted gift from the 3M Foundation awarded to Prof. Cécile Chazot (3M STEM and Skilled Trades Program).

The authors are grateful to Sarah Berman and Komal Prasad for helpful discussions, encouragement, and moral support throughout this work.

## Author Declarations

The authors have no conflicts to disclose.

A.D.: conceptualization, methodology, software, validation, formal analysis, investigation, data curation, writing – original draft, writing – review & editing, visualization

A.M.: software, investigation, data curation, writing – review & editing

D.B.: conceptualization, resources, writing – review & editing, supervision, project administration, funding acquisition

**Data Availability Statement**

The data that support the findings of this study are available from the corresponding authors upon reasonable request.